\documentstyle[twocolumn,aps]{revtex}

\begin{document}

\draft

\title{Transient topological objects in high energy collisions. }
\author{ A. Makhlin   }
\address{Department of Physics and Astronomy, Wayne State University
Detroit, MI 48202}
\date{September 20, 1999}
\maketitle
\begin{abstract}

I study the topology of quantum fluctuations which take place at the
earliest stage of high-energy processes. A new exact solution of Yang-Mills
equations with fractional topological charge and carrying a single color is
found.

\end{abstract}


~{\bf 1}. The problem of initial data for ultra-relativistic heavy ion
collisions has been a sore subject for more than a decade. The roots of the
problem penetrate deeply into the least explored areas of QCD like the nature
of the QCD vacuum and hadronic structure. The parton picture of a nucleus
completely disregards the properties of the vacuum and only partially respects
the hadron structure by replacing the true bounded QCD state with an artificial
flux of free quarks and gluons.  For heavy ion collisions, the evolution of
parton distributions is very different from the evolution in the simplest
processes like {\em ep}-DIS or high-energy {\em pp}-collisions. In a series of
papers \cite{QFK,QGD,TEV} we have studied the causal character of the QCD
evolution and found that high-energy processes explore all possible quantum
fluctuations that may develop {\em before the collision} and are consistent
with a given inclusive probe. All these fields propagate forward in time and
collapse at the vertex of interaction with the probe. These fluctuations are 
the snapshots of nuclei frozen by the measurement, and they cannot be arbitrary
since the emerging final-state configurations must be consistent with all
interactions that are effective on the time-scale of the emission process. In
other words, we have proved that the QGP as the final state can be created only
in a single quantum transition, as an ensemble of collective modes of expanding
matter. The scale of the entire evolution process  appeared to be set by the
physical properties of the final state.

Besides a fair treatment of the final state-interactions, the theory has to
rely on  a realistic picture of the initial state. It must allow one to treat
nuclei as Lorentz-contracted finite-size objects. It has to account for those
interactions that keep nuclei intact before the collisions and are
responsible for the coherence of the nuclei wave function. It is commonly
accepted that the effective interactions responsible for the existence of
hadrons are due to topological fluctuations in the QCD vacuum, i.e.,
instantons. These exist only in Euclidean space, and only spherically
symmetric field configurations are used in the theory that describe stable
hadrons. They cannot be directly transferred to Minkowsky space. 

The solutions of the QCD field equations with  non-trivial topology, by their
design, rely heavily on the identification the directions in physical
(geometric) space with the directions of the internal (tangent) color space
so that the local rotations of the geometric coordinates can be compensated
by the gauge transformation in  color space. It is impossible to match the
$SU(2)\times SU(2)\approx O(4)$ local color group with the Lorentz group,
$SL(2,C)\approx O(1,3)$, since the first one is a compact Lie group, the
second is not. Which of them should be given priority, and what properties
have to be sacrificed in order to identify the angular coordinates of
internal color and external physical spaces? An empirical answer to this
question is already known. One must  use the Euclidean metric and construct
the  self-dual solutions of the Yang-Mills equations (instantons). Acting in
this way, we indeed achieve remarkable successes in the description of many
properties of {\em stable} hadrons \cite{Shuryak}.  These successes are not
by chance and should be considered as important physical inputs. However,
this theory is incapable of describing moving  hadrons. 
Motion is possible only in the Minkowsky world (where no
self-dual fields exist).  Therefore, we may ask if this commonly used
prescription is sufficiently motivated physically? It is perfectly clear that
would such a motivation exist, it can be only due to the nature of the
measurement process: as viewed from Minkowsky world of moving stable hadrons,
the Euclidean calculations provide an effective theory in the rest frame of a
hadron. If a precise resolution of the color field coordinate takes place,
then (since the moment of the measurement) the Euclidean picture is no longer
valid .

~{\bf 2}. In this note, I show that the solutions of the Yang-Mills
equations, which interpolate between the Euclidean and Minkowsky worlds, do
exist. Such an interpolation becomes possible because two regimes are
separated by the light-cone of the point-like interaction. In the Euclidean
domain (before the interaction) the transient topological object has  finite
action and a  fractional winding number. The fields of this object evolve in
Euclidean proper time $\tau$, and after collapse at the time $\tau=0$, they
can be continued as the waves propagating in Minkowsky space. \footnote{
According to English orthography, the suffix {\em -on} in the name of this
object (would it deserves a name) seems unavoidable. I would suggest {\em
ephemeron} (rater than  {\em transiton}) in order not to create an image of a
particle and emphasize an ephemeral nature of this field configuration.}
Constructing the Euclidean part of the solution, I map the $[SU(2)\times
SU(2)]_{color}$ on $[O(4)]_{space}$ and require that  the spin connections of
the metric and  the gauge field potential (both taken in the same group
representation) {\em must be identical}. Acting in this way, I give priority
to the geometry of the gauge group. Moreover, I insist that, before the
interaction has resolved the color field on a sufficiently short scale, the
space-time metric is defined by the internal Euclidean geometry of the gauge
group.

The only tool which is capable of coping with this view of the relationship
between the internal color dynamics and the geometry is the so-called tetrad
formalism (see, e.g.,\cite{Fock,Witten}). Indeed, the vector and spinor fields
are essentially defined in the tangent space.  In a tetrad basis, components of
any tensor (e.g  $A^{\alpha}(x)$, $\gamma^\alpha$) become scalars with respect
to a general coordinate transformations and behave like Lorentz tensors under
the local Lorentz group transformations.  The usual tensors are then given by
the tetrad decomposition, $A^{\mu}(x)=e^{\mu}_{~\alpha}(x)A^\alpha(x)$,
$\gamma^{\mu}(x)=e^{\mu}_{~\alpha}(x)\gamma^\alpha$, etc. The covariant
derivative of the tetrad vector includes two connections (gauge fields). One of
them,  $\Gamma^{\lambda}_{~\mu\nu}(x)$, is the gauge  field which provides 
covariance with respect to the general transformation of coordinates. The
second gauge field,  the spin connection  $\omega_{\mu}^{~\alpha\beta}(x)$,
provides covariance with respect to the local Lorentz rotation.

Let $x^\mu =(\tau,r,\phi,\eta)$ be the contravariant components of the
curvilinear coordinates that cover the past of the hyperplane $t=0,~z=0$ of the
interaction,
\begin{eqnarray}
x^0=-\tau\cosh\eta~,~~~x^3=-\tau\sinh\eta~,\nonumber\\
x^1= r\cos\phi~,~~~x^2=r\sin\phi~,
\label{eq:E2.0}\end{eqnarray}
where $x^\alpha =(t,x,y,z)\equiv(x^0,x^1,x^2,x^3)$ are
those of the flat Minkowsky space. [In order to cover the full Minkowsky space,
we have to build a chart with similar parameterization in each of four domains
separated by the hypersurfaces, $\tau^2=t^2-z^2= 0$. The domains $\tau^2>0$ can
be split further into two parts by the light cone $\tau^2 = r^2$.]
For the coordinates (\ref{eq:E2.0}),  the four tetrad vectors
$e^{\alpha}_{~\mu}$ form a matrix
\begin{eqnarray}
e^{\alpha}_{~\mu}={\rm diag}(1,1,r,\tau)~.
\label{eq:E2.1}\end{eqnarray}
These vectors correctly reproduce the curvilinear metric ${\rm g}_{\mu\nu}$ 
and the flat Minkowsky metric $g_{\alpha\beta}$, {\em i.e.},
\begin{eqnarray}
{\rm g}_{\mu\nu}=g_{\alpha\beta}e^{\alpha}_{~\mu}e^{\beta}_{~\nu}
={\rm diag}[-1,1,r^2,\tau^2]~,\nonumber\\
g^{\alpha\beta}= {\rm g}^{\mu\nu} e^{\alpha}_{~\mu}e^{\beta}_{~\nu}=
{\rm diag}[-1,1,1,1]~.
\label{eq:E2.2}\end{eqnarray}
The spin connection can be found from the condition that the covariant
derivatives of the tetrad vectors equal zero \cite{Witten}, i.e.,
\begin{eqnarray}
\omega^{~\alpha\beta}_{\mu}=[\Gamma^{\lambda}_{~\mu\nu}
e^{\alpha}_{~\lambda}-
\partial_\mu e^{\alpha}_{~\nu}]e^{\beta~\nu}~.\label{eq:E2.3}
\end{eqnarray}
Indeed, the tetrad vector $e^{\alpha}_{~\mu}$ is the coordinate vector and the
Lorentz vector at the same time. (The Lorentz index $\alpha$ and the coordinate
index $\mu$ are moved up and down by the local Minkowsky metric tensor
$g_{\alpha\beta}$ and the global metric tensor ${\rm g}_{\mu\nu}$,
respectively.) The only non-vanishing components of the connections are
\begin{eqnarray}
\Gamma^{\cdot}_{\eta\eta\tau} =
-\Gamma^{\cdot}_{\tau\eta\eta}=-\tau,~ \Gamma^{\cdot}_{\phi\phi r} =
-\Gamma^{\cdot}_{r\phi\phi}=-r~,\nonumber\\
\omega_{\eta}^{~30}=-\omega_{\eta}^{~03}=1,~\omega_{\phi}^{~12}
=-\omega_{\phi}^{~21}=-1~.
\label{eq:E2.4}\end{eqnarray}

In the tetrad formalism,  the transition to the Euclidean space is easily done 
by making the time-like tetrad vector $e^{0}_{~\mu}$ imaginary,
$e^{0}_{~\mu}\to (e^{0}_{~\mu})_E= (i,0,0,0),~~(e_{0}^{~\mu})_{E}=
(-i,0,0,0)$. Then Eqs.~(\ref{eq:E2.2}) take the form
\begin{eqnarray}
{\rm g}_{\mu\nu}=
g_{\alpha\beta}(e^{\alpha}_{~\mu})_{E}(e^{\beta}_{~\nu})_{E}
={\rm diag}[\pm 1,1,r^2,\tau^2]~,\nonumber\\
g^{\alpha\beta}= {\rm g}^{\mu\nu}
(e^{\alpha}_{~\mu})_{E}(e^{\beta}_{~\nu})_{E}=
{\rm diag}[\mp 1,1,1,1]~.
\label{eq:E2.5}\end{eqnarray}
This formal step also leads to a set of standard prescriptions for the
transition to the Euclidean version of the field theory:
$A^\tau_E=(e_{0}^{~\tau})_{E}A^0 =-iA^0$. The same rule holds for the
spin connection, $(\omega_{\mu}^{~03})_{M}~\to~
(\omega_{\mu}^{~03})_{E}= -i (\omega_{\mu}^{~03})_{M}$. These formulae
indicate that we perform a transition to an {\em imaginary proper time}
$\tau$. The choice between the signs in Eq.~(\ref{eq:E2.5})  is a
subject of an independent analysis.

~{\bf 3}. Assuming the Euclidean long-distance behavior, we employ the metric
\begin{eqnarray}
ds^2=d\tau^2 +dr^2+r^2 d\phi^2 +\tau^2 d\eta^2
\label{eq:E3.0}\end{eqnarray}
with the only non-vanishing
components of the spin connection being $\omega_{\eta}^{~03}=-1$, and
$\omega_{\phi}^{~12}=-1$. [Christoffel symbols $\Gamma$ remain the same as in
Eq.(\ref{eq:E2.4}). Overall, we have four domains with the same Euclidean
metric, which explains the result (\ref{eq:E4.10},\ref{eq:E4.13}).] 
Introducing the (iso)vector representation
$A^{~\alpha\beta}_{\mu}$ of the gauge field of the $O(4)$ group, and insisting
on a one-to-one mapping of the color and space directions, we require that
\begin{eqnarray}
 A^{~\alpha\beta}_{\mu}=h~\omega^{~\alpha\beta}_{\mu}~,
\label{eq:E3.1}\end{eqnarray}
where the factor $h$ is an arbitrary real number which defines the relationship
between the cyclic components of space-time and color coordinates. 
 The gauge fields of $O(4)$ have two projections on its two $SU(2)$-subgroups,
\begin{eqnarray}
(A^{a}_{~\mu})_{\pm}={1\over 2}~\eta^{a\alpha\beta}_{\pm}A^{~\alpha\beta}_{\mu}
 = \pm A^{0\alpha}_{\mu} +{1\over 2}\epsilon^{a\alpha\beta} 
 A^{~\alpha\beta}_{\mu}~,
\label{eq:E3.2}\end{eqnarray}
where $\eta^{a\alpha\beta}_{\pm}$ are the 't Hooft symbols \cite{'t Hooft},
and the subscripts $(\pm)$ denote two chiral projections. Thus, we have
\begin{eqnarray}
 (A^{3}_{~\eta})_{\pm}=\mp~h~\omega^{~03}_{\eta}=\pm h,~~
 (A^{3}_{~\phi})_{\pm}=h\omega^{~12}_{\phi}= - h,
\label{eq:E3.3}\end{eqnarray}
which is compatible with the gauge condition $A^\tau =0$  that we adopt
for both the Euclidean and the Lorentz regimes of the process.
One can easily find a representation for this
potential which manifests its pure gauge origin,
\begin{eqnarray}
A_{\mu}(x)=(1/2)A^{a}_{\mu}(x)\sigma^a=S\partial_\mu S^{-1}~.
\label{eq:E3.4}\end{eqnarray}
Using the decomposition, $S=iu_0\bbox{1}+u_a\bbox{\sigma}^a$, and
$S^{-1}=-iu_0\bbox{1}+u_a\bbox{\sigma}^a$ we arrive at
\begin{eqnarray}
A_{\mu}(x)=(1/2)A^{c}_{\mu}(x)\bbox{\sigma}^c \nonumber\\
=-(\epsilon^{abc}u_a\partial_\mu u_b + u_0\partial_\mu u_c-
u_c\partial_\mu u_0)\bbox{\sigma}^c~.
\label{eq:E3.6}\end{eqnarray}
By comparison with (\ref{eq:E3.3}), and accounting for the unitarity,
$SS^{-1}=1$, we obtain a system of equations,
\begin{eqnarray}
 -2(u_1\partial_\eta u_2 -
u_2\partial_\eta u_1 + u_0\partial_\eta u_3 -
u_3\partial_\eta u_0)=\pm h~,\nonumber \\
-2(u_1\partial_\phi u_2 -
u_2\partial_\phi u_1 + u_0\partial_\phi u_3 -
u_3\partial_\phi u_0)= - h~,\nonumber\\
u_0^2 + u_a^2=1~~, \label{eq:E3.7}
\end{eqnarray}
which has a solution
\begin{eqnarray}
(u_0)_{\pm}= \mp 2^{-1/2} \cos h\eta,~~
(u_3)_{\pm}=  2^{-1/2} \sin h\eta~, \nonumber  \\
(u_1)_{\pm}=  2^{-1/2} \cos h\phi,~~
(u_2)_{\pm}=  2^{-1/2} \sin h\phi~.
\label{eq:E3.8}\end{eqnarray}

A non-trivial solution of the Yang-Mills equations is looked for in 
the form
\begin{eqnarray}
(A^{3}_{~\eta})_{\pm}= E(\tau,r)~iS\partial_\eta S^{-1}=
\mp E(\tau,r) \to \pm 1,\nonumber  \\
(A^{3}_{~\phi})_{\pm}= \Phi (\tau,r)~iS\partial_\phi S^{-1}=
 -\Phi(\tau,r) \to -1,
\label{eq:E3.9}\end{eqnarray}
where the arrows point to the values of potentials at $\tau\to\infty$, 
where the
field must approach a pure gauge. (For now, we assume that $h=1$.) Since the
field has only one color component, the commutator in the definition,
$F_{\mu\nu}=\partial_\mu A_\nu - \partial_\nu A_\mu -[A_\mu ,A_\nu]$, vanishes
and the components of the field tensor are the same as 
in the Abelian case where
\begin{eqnarray}
F_{\tau\eta}=\mp\partial_\tau E,~~F_{\tau \phi}=-\partial_\tau \Phi,~~
F_{\tau r}=0~, \nonumber \\
F_{r\eta}=\mp\partial_r E,~~F_{r\phi}=-\partial_r \Phi,~~
F_{\eta\phi}=0~.
\label{eq:E3.10}\end{eqnarray}
The condition for the (anti)self-duality of the field tensor
$F_{\mu\nu}$ reads as
\begin{eqnarray}
\overstar{F}_{\mu\lambda}\equiv {\rm g}_{\mu\nu} {\rm g}_{\lambda\sigma}
{\epsilon^{\nu\sigma\rho\kappa}\over 2\sqrt{\rm g}} F_{\rho\kappa}
= \pm F_{\mu\lambda}~.
\label{eq:E3.11}\end{eqnarray}
Note that the definition of the dual tensor is different from the
familiar definition in flat space. This modification is obvious.
Indeed, the co- and contravariant tensor components are
even of different dimensions. The requirement of the self-duality of the
field (\ref{eq:E3.9}) yields a system of equations,
\begin{eqnarray}
{1\over r}{\partial \Phi \over \partial r}=
{1\over \tau}{\partial E \over \partial \tau},~~~
\tau{\partial \Phi \over \partial \tau} =
- r{\partial E \over \partial r}~.
\label{eq:E3.12}\end{eqnarray}
Two conditions of the self-consistency for this system are
\begin{eqnarray}
\partial_r^2 \Phi-{1\over r}\partial_r\Phi=
-(\partial_\tau^2\Phi +{1\over \tau}\partial_\tau\Phi)~,\nonumber \\
\partial_r^2 E +{1\over r}\partial_r E=
-(\partial_\tau^2 E - {1\over \tau}\partial_\tau E)~.
\label{eq:E3.13}\end{eqnarray}
This system is easily solved by separation of variables. The
solution which obeys the original system (\ref{eq:E3.12}), the condition
for finiteness, the boundary condition of pure gauge at $\tau\to\infty$, 
and the condition that $A_\eta\to 0$ at $\tau\to 0$, is as follows,
\begin{eqnarray}
\mp A^3_\eta=
E(\tau,r) = -\lambda\tau K_1(\lambda\tau)J_0(\lambda r)+1,\nonumber\\
- A^3_\phi=
\Phi (\tau,r) = \lambda r J_1(\lambda r) K_0(\lambda\tau) + 1.
\label{eq:E3.14}\end{eqnarray}
From these expressions for the potentials, we may easily find
the field strengths of the ephemeron,
\begin{eqnarray}
\mp\tau^{-1}F^3_{\tau\eta}=-r^{-1}F^3_{r\phi}
=\lambda^2 K_0(\lambda\tau)J_0(\lambda r)~,\nonumber\\
\mp r^{-1} F^3_{\tau\phi}=-\tau^{-1}F^3_{r\eta}
=\lambda^2  J_1(\lambda r) K_1(\lambda\tau)~.
\label{eq:E3.15}\end{eqnarray}

The geometry of this field is noteworthy. It has the symmetry of a
torus. The magnetic field has two components, one along the torus
pipe, and the second winding around the pipe. This is a well known
configuration of a toroidal magnetic trap. Since $r<\tau$, both radii
of the torus  get smaller when $\tau\to 0$; the torus collapses at
$\tau=0$.  The electric fields has two similar components which are
created in accordance with  the Abelian induction law. Every pipe is 
mono-colored. The parameter $\rho=\lambda^{-1}$ clearly plays the role of the
``size'' of the ephemeron.

~{\bf 4}.  Starting from the fields given by the Eqs.~(\ref{eq:E3.10})
we may find the Euclidean action of the ephemeron:
\begin{eqnarray}
 S_E = -{1\over 4 g^2}\int d^4x \sqrt{\rm g}~ {\rm g}^{\mu\rho}
 {\rm g}^{\nu\sigma} F^a_{\mu\nu} F^a_{\rho\sigma} \nonumber\\
 =-{2\pi^2\over  g^2}\int_{0}^{\infty}\tau
 d\tau\int_{0}^{\tau}{dr\over r}
 \bigg[\bigg({\partial\Phi\over\partial\tau}\bigg)^2+
 \bigg({\partial\Phi\over\partial r}\bigg)^2\bigg]~.
\label{eq:E4.1}\end{eqnarray}

In the same way, we compute the topological charge
\begin{eqnarray}
 Q= {1\over 32 \pi^2}\int d^4x \sqrt{\rm g}
 ~{\epsilon^{\mu\nu\rho\sigma}\over 2\sqrt{\rm g}}
 ~ F^a_{\mu\nu} F^a_{\rho\sigma} \nonumber\\
 =\pm {1\over 2}\int_{0}^{\infty}\tau
 d\tau\int_{0}^{\tau}{dr\over r}
 \bigg[\bigg({\partial\Phi\over\partial\tau}\bigg)^2+
 \bigg({\partial\Phi\over\partial r}\bigg)^2\bigg]~.
 \label{eq:E4.2}\end{eqnarray}
Thus, we have reproduced a standard relation, between
the instanton action and its winding number,
\begin{eqnarray}
S_E~=~-~{8\pi^2 \over g^2}~|Q|~.
\label{eq:E4.3}\end{eqnarray}
We now have to find the winding number $Q$.

We shall do it using the representation of topological
charge via the divergence of the Chern-Simons current,
\begin{eqnarray}
Q={1\over 4\pi^2} \oint d\sigma_\mu K^\mu,
\label{eq:E4.4}\end{eqnarray}
where
\begin{eqnarray}
K^\mu ={1\over 4} \epsilon^{\mu\nu\rho\sigma}
\big[A_\nu^a\partial_\rho A_\sigma^a +
{g\over 3}\epsilon_{abc}A_\nu^a A_\rho^b A_\sigma^c \big]~.
\label{eq:E4.5}\end{eqnarray}
The second term (usually the major one) identically
vanishes since the ephemeron field has only one color component.
In our geometry, only two components of $K^\mu$ survive,
\begin{eqnarray}
 K^\tau = {\pm 1\over 4}\bigg[E{\partial \Phi\over \partial r}
             - \Phi{\partial E\over \partial r}\bigg]=
        {\pm 1\over 8}\bigg[{r\over\tau}{\partial E^2 \over \partial\tau}
        + {\tau\over r} {\partial \Phi^2\over \partial \tau}\bigg]
\label{eq:E4.6}\end{eqnarray}
and
\begin{eqnarray}
 K^r = {\mp 1\over 4}\bigg[E{\partial \Phi\over \partial \tau}
             - \Phi{\partial E\over \partial r}\bigg]=
        {\mp 1\over 8}\bigg[{r\over\tau}{\partial E^2 \over \partial r}
        + {\tau\over r} {\partial \Phi^2\over \partial r}\bigg].
\label{eq:E4.7}\end{eqnarray}
Correspondingly, the total flux of the vector $K^\mu$ over a closed
surface can be split into the sum of integrals over three
surfaces, $\tau=\infty$, $r=0$, and $\tau=r$,
\begin{eqnarray}
Q=\int d\tau dr \theta(\tau -r)[\partial_\tau K^\tau+\partial_r K^r]\nonumber\\
=\int_0^\infty dr[K^\tau(\infty,r)-K^\tau(r,r)] \nonumber\\
+\int_0^\infty d\tau[K^r(\tau,\tau)-K^r(\tau,0)] .
\label{eq:E4.8}\end{eqnarray}
(The factor $4\pi^2$ has come from two angular integrations.)
Straightforward computation of the integrals lead to the following
expressions,
\begin{eqnarray}
\int_0^\infty K^\tau(\infty,r)dr=0,
~~-\int_0^\infty K^r(\tau,0)d\tau={1\over 4}~,\nonumber\\
\int_0^\infty [K^r(\tau,\tau)-K^\tau(\tau,\tau)]d\tau=-{1\over 8}~~.~~~~
\label{eq:E4.10}\end{eqnarray}
Eventually, we find
\begin{eqnarray}
Q= {1\over 4\pi^2} \oint d\sigma_\mu K^\mu={1\over 8}~.
\label{eq:E4.13}\end{eqnarray}
The transient topological gluon configuration carries a fractional topological
charge of 1/8 in that part of the Euclidean space which is an image of the 
interior of the  past light cone of the interaction vertex. In a full chart,
we have several domains with similar Euclidean picture, and the sum over all 
of them gives $Q_{\rm tot}=1$.

~{\bf 5.} The ephemeron field is defined by four parameters. Two of them, $x_0$
and $y_0$, are obvious and correspond to the preserved translation symmetry in
the $xy$-plane. To include them explicitly, we must read $r$ as
$\sqrt{(x-x_0)^2+(y-y_0)^2}$. Next, we have the ephemeron ``radius''
$\rho=\lambda^{-1}$. The fourth parameter is the scale factor $h$, which  has
been introduced in Eq.~(\ref{eq:E3.1}) an dropped from the explicit
calculations after Eq.~(\ref{eq:E3.8}).  In fact, the presence of scale factor
$h$ among the  parameters of ephemeron can be recovered from the observation
that equations (\ref{eq:E3.12}) are homogeneous (and thus admit an arbitrary
scale factor) and the fact that solutions  for the vector potential then have a
pure gauge asymptote given by Eqs.~(\ref{eq:E3.4})-(\ref{eq:E3.8}). With $h$
explicitly retained, the topological charge is $Q=h^2/8$ and thus becomes a
continuous parameter. This is in contrast with the case of spherically
symmetric BPST instantons \cite{BPST} which can have only integer charges.
Equations (\ref{eq:E3.6}) and (\ref{eq:E3.8}) explain the origin of the
non-trivial topology of the ephemeron. The solutions with different values of
$h$ correspond to different field configurations whose asymptotes  at
$\tau\to\infty$ are the different gauge transforms of the classical vacuum with
$A(x)=0$.

The most impressive feature of the ephemeron is that it is one-dimensional both
in physical and color spaces. This is not so surprising from a mathematical
point of view since we map the two spaces onto one another. In physical space,
the full spherical symmetry is corrupted by the interaction and only two
rotations in the $tz$- and the $xy$-planes survive as an actual symmetry. The
high-precision measurement of the coordinate inside an object formed by the
strong interaction necessarily resolves a mono-colored field pattern.
Topologically, the ephemeron is a collapsing ring before the interaction, 
and an opened expanding string after it.

Existence of the transient topological field configurations poses several
important questions. 

(i) How does the toroidal geometry of the ephemeron affect
the distribution of the gluons produced in high-energy collisions? Is the
proper field of the resolved color charge  Coulomb-like or does it carry
some remnants of the  twisted geometry of the ephemeron fields. 

(ii) The geometry of the electric and magnetic field of the ephemeron implies
strong spin polarization effects. All known evidences of  $P$- and
$CP$-violations come from the study of various decays which are genuinely
non-stationary processes. What is the role of the transient topological
configurations in these decays? Formally, the ephemerons must
be included into the path-integral representation of the point-to-point
correlators of hadronic currents on equal footing with the BPST instantons.

Finally, the border between perturbative and non-perturbative QCD is clearly 
defined by their relation to the non-trivial topological properties of the QCD
vacuum. The very existence of the ephemeron solutions of the Yang-Mills
equations undoubtedly indicates that there is a hope to bridge the gap.

This work was supported by the U.S. Department of Energy under
Contract  No. DE--FG02--94ER40831.

\end{document}